\documentclass[twocolumn,amsmath,amssymb,prb]{revtex4}
\usepackage{amsmath}
\usepackage{graphicx}% Include figure files
\usepackage{graphicx}
\begin{document}
\title{
Response of a Hodgkin-Huxley neuron to a high-frequency input
}

\author{L.~S.~Borkowski}

\affiliation{Faculty of Physics, Adam Mickiewicz University,
Umultowska 85, 61-614 Poznan, Poland}

%\ead{lsb@man.poznan.pl}

\begin{abstract}
We study the response of a Hodgkin-Huxley neuron
stimulated by a periodic sequence
of conductance pulses arriving through the synapse
in the high frequency regime.
In addition to the usual excitation threshold
there is a smooth crossover from the firing to the silent regime
for increasing pulse amplitude $g_{syn}$.
The amplitude of the voltage spikes decreases approximately
linearly with~$g_{syn}$.

In some regions of parameter space the response
is irregular, probably chaotic.
In the chaotic regime between
the mode-locked regions 3:1 and 2:1
near the lower excitation threshold
the output interspike interval histogram (ISIH)
undergoes a sharp transition.
If the driving period is below the critical value, $T_i < T^*$,
the output histogram contains only odd multiples of $T_i$.
For $T_i > T^*$ even multiples of $T_i$ also
appear in the histogram, starting from the largest values. 
Near $T^*$ the ISIH scales logarithmically
on both sides of the transition.
The coefficient of variation of ISIH has a cusp singularity
at $T^*$. The average response period has a maximum
slightly above $T^*$.
Near the excitation threshold in the chaotic regime
the average firing rate
rises sublinearly from frequencies of order 1 Hz.

\end{abstract}
%\pacs{87.19.ll,87.19.ln}
%%%%%%%%%%%%%%%%%%%%%%%%%%%%%%%%%%
\maketitle

\section{Introduction}

Biological neurons transmit information
in the form of sharp spikes of potential difference
across the lipid bilayer forming the wall of the nerve
cell. This feature of the cell's reaction to input
signals is remarkably consistent in different organisms
and different types of neurons. The action potential spikes
are assumed to be the principal carrier of information.
The early view that information is transmitted via
rate coding has evolved. It is now recognized that
also the spike time coding is used in neural systems\cite{Sejnowski95,Ferster95}.
While the precise coding recipe is unknown it is clear
that the knowledge of the response of various types of neurons
to different stimuli is fundamental to formulating
the theory of information transfer in the neural system.

Our understanding of conductance-based models of neurons
is largely based on the Hodgkin-Huxley (HH)
model originally formulated to describe the dynamics of the membrane potential
of the squid giant axon\cite{HH1952}.
The detailed voltage-clamp measurements
of the voltage-gated potassium and sodium ion currents led
to revisions of the HH model.
The modifications required to achieve better agreement
with experiments were reviewed by Clay\cite{Clay2005}.
Studies of single neurons and neuronal networks often employ
simplified models, such as integrate-and-fire
and FitzHugh-Nagumo (FHN) models\cite{Gerstner1995,Bressloff1998}.
It is believed that the two-dimensional flow models
such as FHN reproduce qualitatively the behavior
of the HH model.
However these simplifications
are not always justifiable
\cite{Rinzel1980,Brown1999,Guckenheimer2002}.
In an interesting analysis of chaos in the HH model
Guckenheimer and Oliva\cite{Guckenheimer2002}
point out that even the concept of a firing threshold
may be more subtle that just a smooth hypersurface
dividing subthreshold and suprathreshold membrane potentials.

Over the years many studies of HH equations were carried out,
including stochastic variations
of various quantities\cite{Leekim99,Mosekilde05,Luccioli06}.
An important question is to what extent
the qualitative properties of neuron response
depend on the functional form
of the input signal. One frequently used form
of input is constant plus a sinusoidal term.
However the physiological signals are more pulse-like.
In a strongly nonlinear system this may lead
to substantial differences in the output.

In the sinusoidally driven HH model
the excitation threshold rises
sharply at large frequencies.
The phase diagram in the frequency-current amplitude plane consists
of three phase locked regions with integer ratio
of the output period to the input period, $\bar{T_o}/T_i$, 1:1, 2:1, and 3:1.
There are also areas of fractional locking and bistable or chaotic
response around these phase-locked states\cite{Lee-kim2006,Chik2001,Che2009}.

It was pointed out that the edges of mode-locked plateaus
have analogies to phase transitions in the equilibrium
statistical mechanics. Two forms of scaling of the average deviation
from perfect mode-locking were found near the edges
of plateaus with constant $p/q$, where $p$ and $q$ are integers,
indicating number of input spikes per number of output
action potentials\cite{Engelbrecht09}.
The scaling has either exponent $1/2$ or is logarithmic.
In this paper we will show that scaling is more common
and appears also near the multimodal transition points.

Here we assume the $\alpha$ form of postsynaptic current,
$I_{syn} \sim t\exp(-t/\tau)$, where $t$ is time from
the onset of the input spike and $\tau$
is the time scale of the synaptic action.
This form is close to experimental observation although
it does not take into account a more complex dynamics
of the ion channel kinetics, usually described
in the Markovian scheme.

The general form of the phase diagram of the Hodgkin-Huxley model
with this input was studied initially in Ref. \cite{Hasegawa2000}.
However many important questions are still to be answered. One of them is
the behavior of the system in the high-frequency limit.
In the following we present the model and show
the main features of high-frequency response.

\section{The model}

The Hodgkin-Huxley neuron subject to periodic conductance pulses
is defined by the following set of equations,\cite{HH1952}

\begin{equation}
\label{HH}
\begin{split}
C dV/dt = - g_{Na} m^3 h(V-V_{Na}) - g_K n^4 (V-V_K)\\
- g_L (V-V_L) +I_{ext}+I_{syn} ,
\end{split}
\end{equation}

\begin{equation}
dm/dt = -(a_m+b_m) m + a_m ,
\end{equation}

\begin{equation}
dh/dt = -(a_h+b_h) h + a_h ,
\end{equation}

\begin{equation}
dn/dt = -(a_n+b_n) n + a_n ,
\end{equation}

\noindent
where

\begin{equation}
\label{am}
a_m = 0.1(V+40) / [1 - e^{-(V+40)/10}]  ,
\end{equation}

\begin{equation}
\label{bm}
b_m = 4 e^{-(V+65)/18} ,
\end{equation}

\begin{equation}
\label{ah}
a_h = 0.07 e^{-(V+65)/20}  ,
\end{equation}

\begin{equation}
\label{bh}
b_h = 1/[1+e^{-(V+35)/10}] ,
\end{equation}

\begin{equation}
\label{an}
a_n = 0.01 (V+55)/[1-e^{-(V+55)/10}]  ,
\end{equation}

\begin{equation}
\label{bn}
b_n = 0.125 e^{-(V+65)/80} .
\end{equation}
In equations (\ref{am})-(\ref{bn}) the voltage is expressed in mV and the rate constants $\alpha$ and $\beta$ are given in $\textrm{ms}^{-1}$.
The reversal potentials of sodium, potasium and leakage channels
are $V_{Na} = 50 \textrm{mV}$, $V_K = -77 \textrm{mV}$,
and $V_L = -54.5 \textrm{mV}$, respectively. 
The corresponding maximum conductances are $g_{Na} = 50 \textrm{mV}$,
$g_K = 36 \textrm{mS/cm}^2$, and $g_L = 0.3 \textrm{mS/cm}^2$.
The capacity of the membrane is $C = 1 \mu \textrm{F/cm}^2$\cite{HH1952}.
 
\begin{figure}[th]
\includegraphics[width=0.48\textwidth]{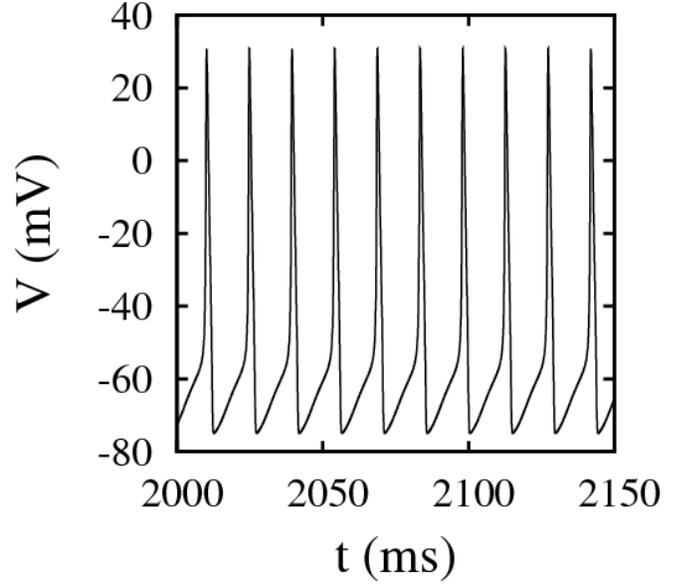}
\caption{Sample voltage trace for a constant input current
$I_{ext} = 10 \mu\textrm{A/cm}^2$.
}
\label{przebieg1}
\end{figure}

The synaptic current $I_{syn}$ is given by the following equation,

\begin{equation}
I_{syn}(t) = g_{syn}\sum_n \alpha(t-t_{in})(V_a-V_{syn}) ,
\end{equation}
where $t_{in}$ denotes the start of the $n^{th}$ pulse, $g_{syn}$ is the conductivity of the synapse, $V_a = 30 mV$
is the maximum potential in the postsynaptic area and
$V_{syn} = -50 mV$ is the reversal potential of the synapse.
The period of the synaptic drive is 
$T_i = t_{in+1}-t_{in}$.
The external current $I_{ext}$ is set to 0, except for a sample
run shown in Fig. \ref{przebieg1}.

\begin{figure}[th]
\includegraphics[width=0.44\textwidth]{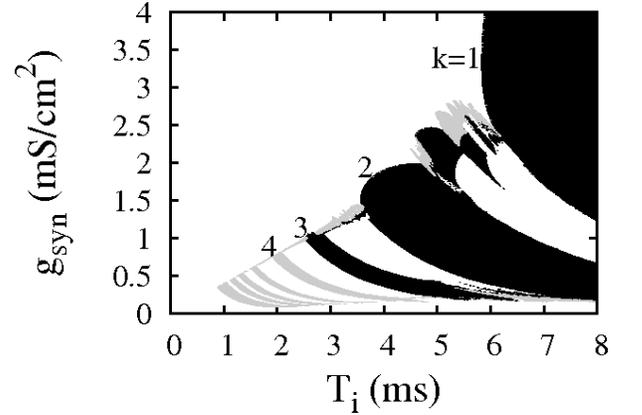}
\caption{The ratio of the average output spiking
rate to the input rate, $k=\bar{T_o}/T_i$. Mode-locked regions with $k=1,2,3$
and $k=4,5,6,7,8$ are shown in black and grey respectively.
Voltage peaks were counted as spikes when $V$ exceeded 0.
For high values of $g_{syn}$ the neuron does not respond.}
\label{phasediagram}
\end{figure}

\begin{figure*}[ht]
\includegraphics[width=.3\textwidth]{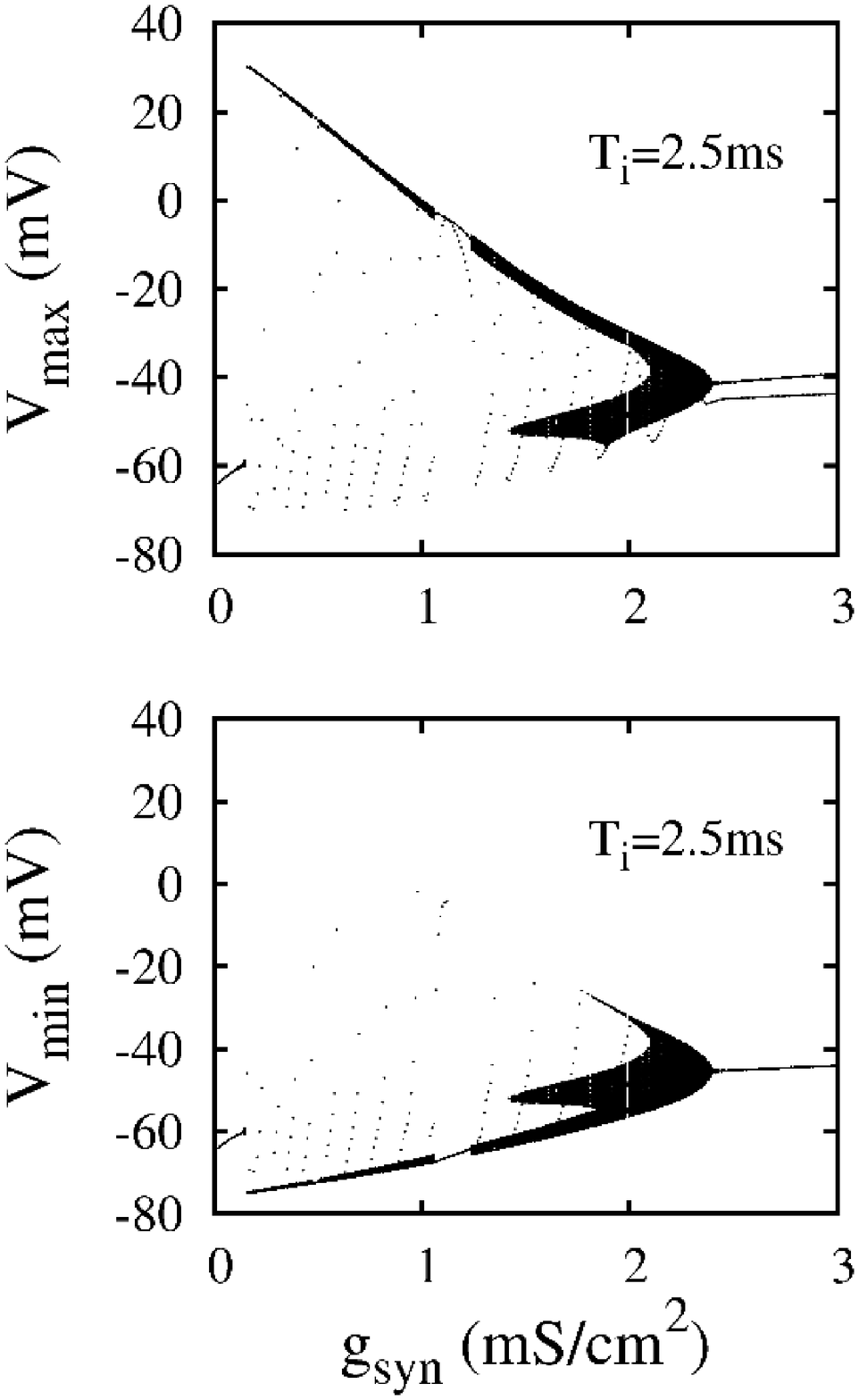}
\includegraphics[width=.3\textwidth]{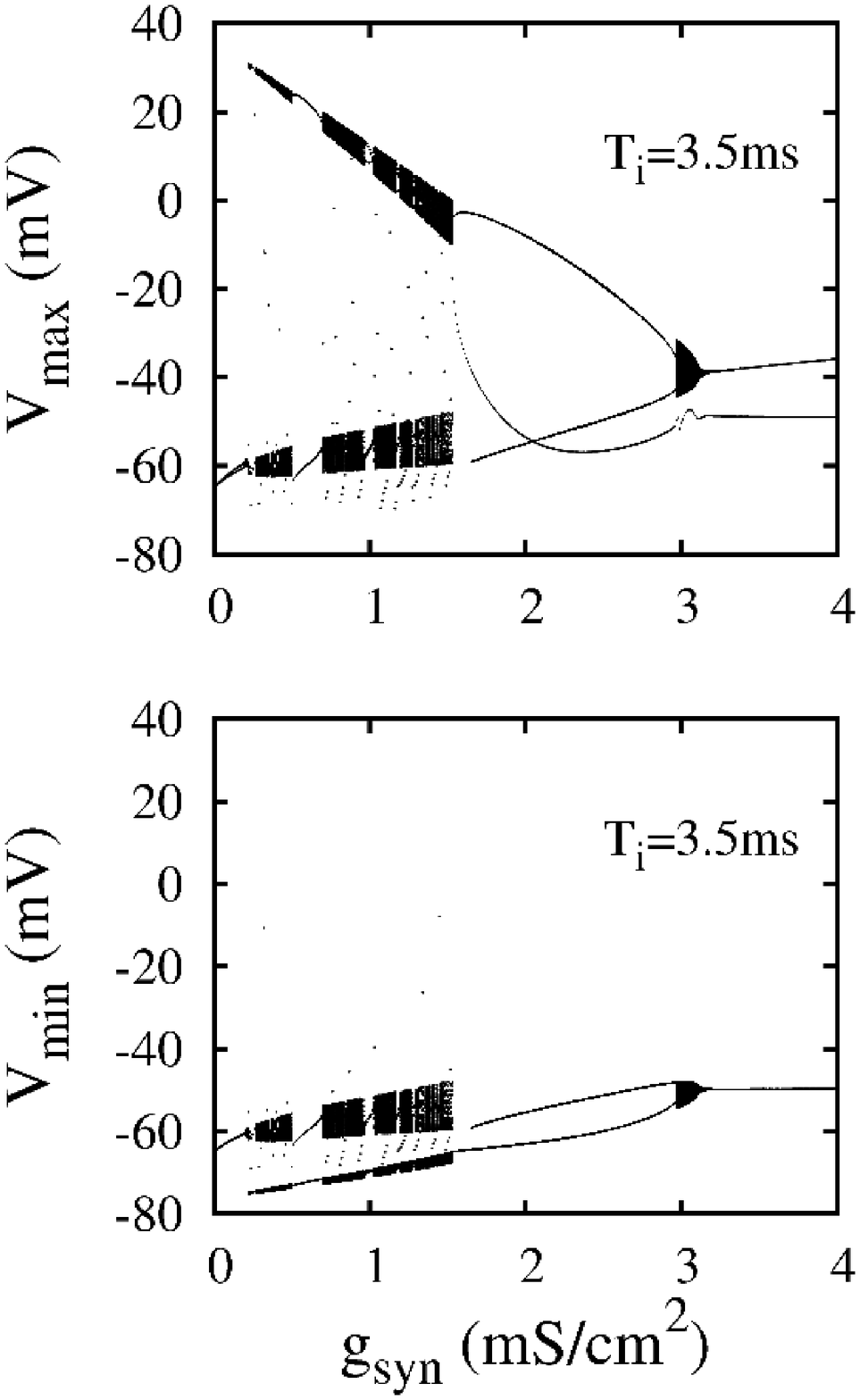}
\includegraphics[width=.3\textwidth]{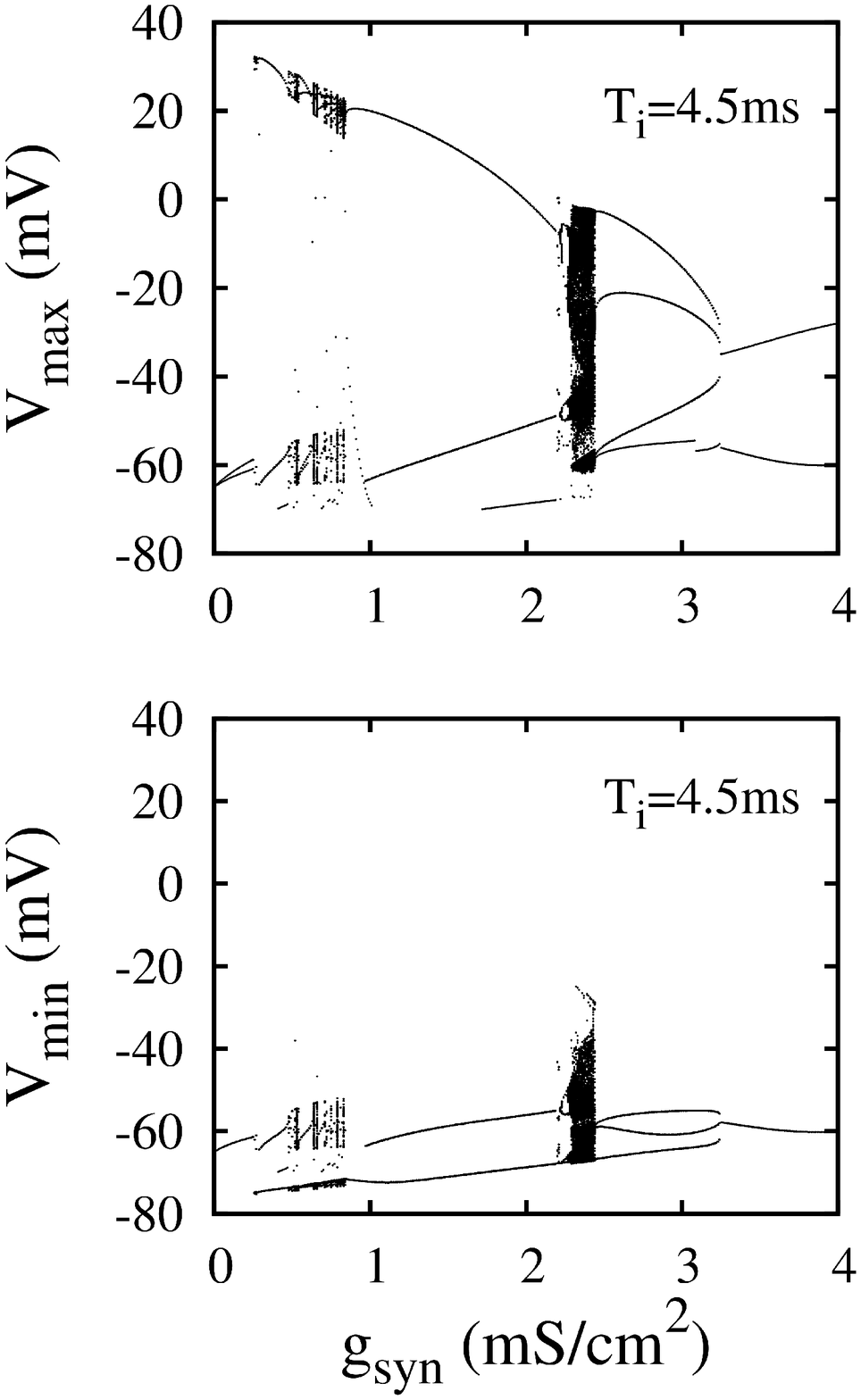}
\caption{Values of maxima and minima of the membrane potential
$V(t)$ as a function of synaptic conductivity $g_{syn}$ for
input spike intervals $T_i = 2.5 \textrm{ms}, 3.5 \textrm{ms}$,
and $4.5 \textrm{ms}$.}
\label{minmax}
\end{figure*}

The time-dependence is given by the function 

\begin{equation}
\label{alfa}
\alpha(t) = (t/\tau) e^{-t/\tau)}\Theta(t) ,
\end{equation}
where $\tau$ is time scale characterizing the dynamics
of the synaptic action and $\Theta (t)$ is the Heaviside step function.
We study the dependence
of the output interspike separation $T_o$
on $T_i$ and $g_{syn}$.

Equations (\ref{HH})-(\ref{bn}) were integrated with
the fourth order Runge-Kutta scheme. The time step was 0.01 ms. For each parameter set the simulation was run for 30 seconds.
Results of the initial three seconds of each data set were
discarded to avoid transient behavior.
In the chaotic regime the data
were obtained from five runs for each value of the horizontal coordinate.
%Fig. \ref{g=0.2} includes results of 100 runs for each $T_i$.

\section{Results}

\begin{figure}[ht]
\includegraphics[width=.45\textwidth]{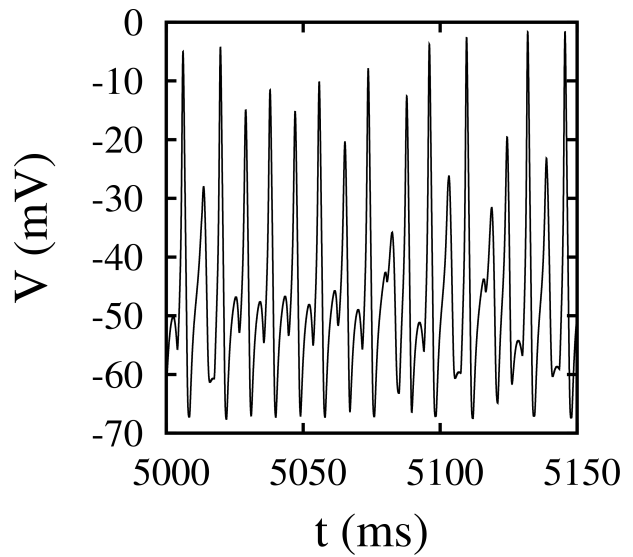}
%\vskip -2.cm
\caption{For high synaptic conductivities the distinction between
action potential and the background oscillations loses its meaning.
This sample was obtained for $T_i=4.5 ms$ and $g_{syn}=2.35
\textrm{mS/cm}^2$.
}
\label{przebieg}
\end{figure}

The average output spiking rate in the form of a color map
as a function of the input period $T_i$
and maximum synaptic conductivity $g_{syn}$
is presented in Fig. \ref{phasediagram}.
The mode-locked regions are shown as areas of uniform color.
For small $T_i$ the total incoming current
is approximately constant with a small modulation,
and the excitation threshold rises linearly with increasing $T_i$,
$g_{syn} \simeq 0.04 T_i ~\textrm{mS/(ms cm}^2\textrm{)}$.
For $g_{syn}$ exceeding approximately
$0.4 T_i ~\textrm{mS/(ms cm}^2\textrm{)}$
the spiking action does not occur.
We can see from Fig. \ref{phasediagram} that
this behavior sets in below $T_i \simeq 6 \textrm{ms}$ .

The obtained phase diagram is qualitatively different
from a response to a sinusoidal input, where
the excitation threshold
diverges as $1/T_i$, for $T_i \rightarrow 0$.
In general we may expect that
the constraint of charge balancing, $\int_t^{t+T_i} Idt=0$,
will have a significant
impact at high frequencies.
For intermediate values of the input period,
$5 \textrm{ms} < T_i < 13 \textrm{ms}$,
the topology of the phase diagram
resembles results obtained with sinusoidal input,
see e.g. Fig. 2 of Ref. \cite{Lee-kim2006}.

Fig. \ref{minmax} shows dependence of minima and maxima
of $V$ on $g_{syn}$ for three input frequencies.
The amplitude of response decreases linearly with increasing $g_{syn}$.
There is no well-defined spiking threshold.
There are intervals of parameter values for which
the response is highly irregular and the values
of maxima and minima of $V$~vary~significantly.

\begin{figure}[ht]
\includegraphics[width=.44\textwidth]{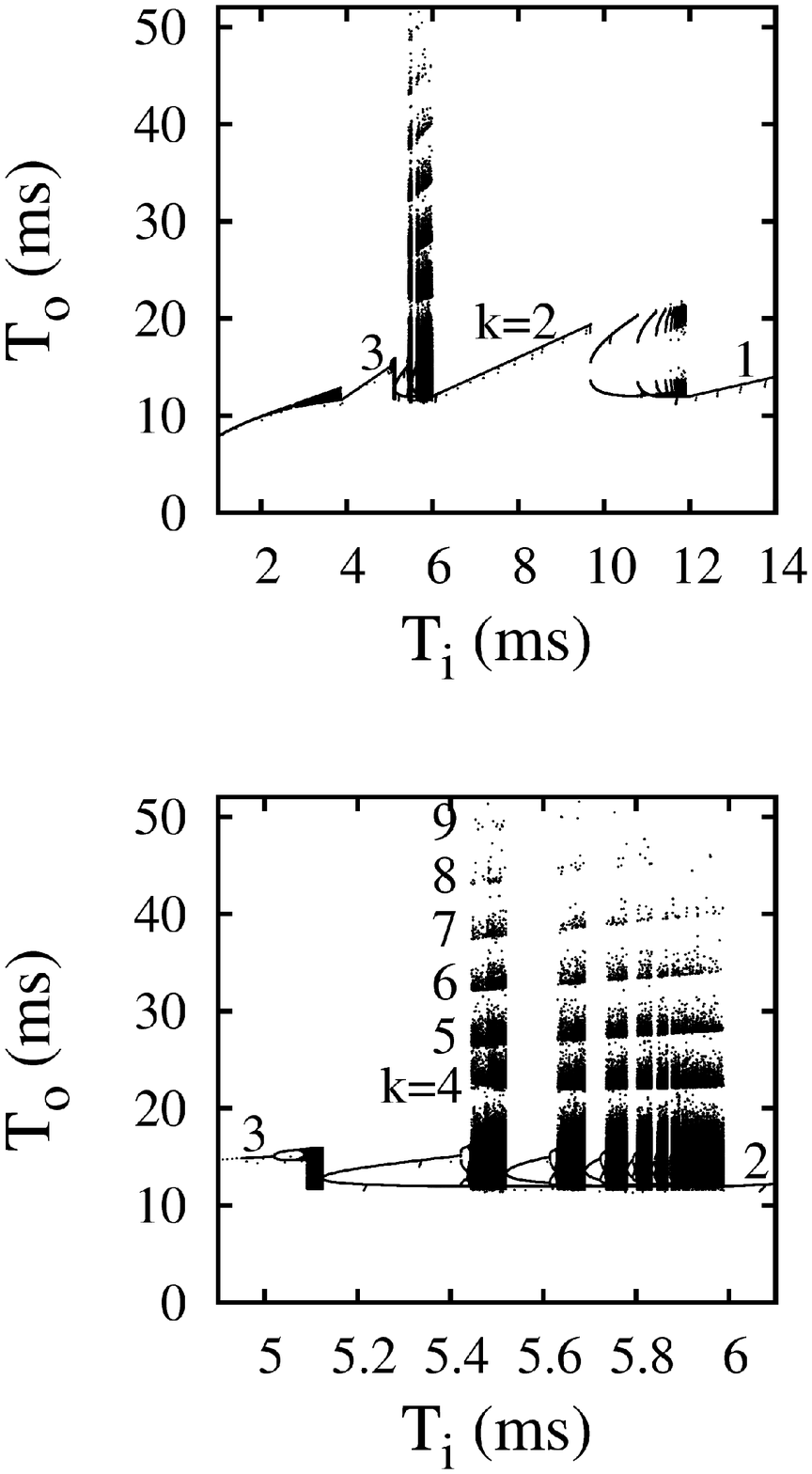}
\caption{
(a)The spectrum of interspike separations of the output signal
as a function of the input period $T_i$
for $g_{syn}=0.4 \textrm{mS/cm}^2$,
(b) Detailed view of the chaotic region between
$T_i = 5 \textrm{ms}$ and $6 \textrm{ms}$.
Each ISI cluster belongs to different $k$, where
$k = 2,3,4,5,...$. The distinction between $k=2$ and $k=3$ is blurred.
}
\label{g04}
\end{figure}

A sample time-dependence of the membrane potential
is shown in Fig. \ref{przebieg}.
The maxima of $V$ span almost
the entire range between $-60 \textrm{mV}$ and $0 \textrm{mV}$.
There is no clear separation of spikes
from the rest of the signal.

Chaotic behavior in the parameter space between
the 3:1 and 2:1 mode-locked regions 
leads to multimodal response. 
The interspike separation for $g_{syn}=0.4 \textrm{mS/cm}^2$
is shown in Fig. \ref{g04}.
For $T_i$ between $5.5 \textrm{ms}$ and $6 \textrm{ms}$
all integer multiples of input $T_i$ with the exception of the lowest one
appear in the output ISIH.

\begin{figure}[ht]
\includegraphics[width=.48\textwidth]{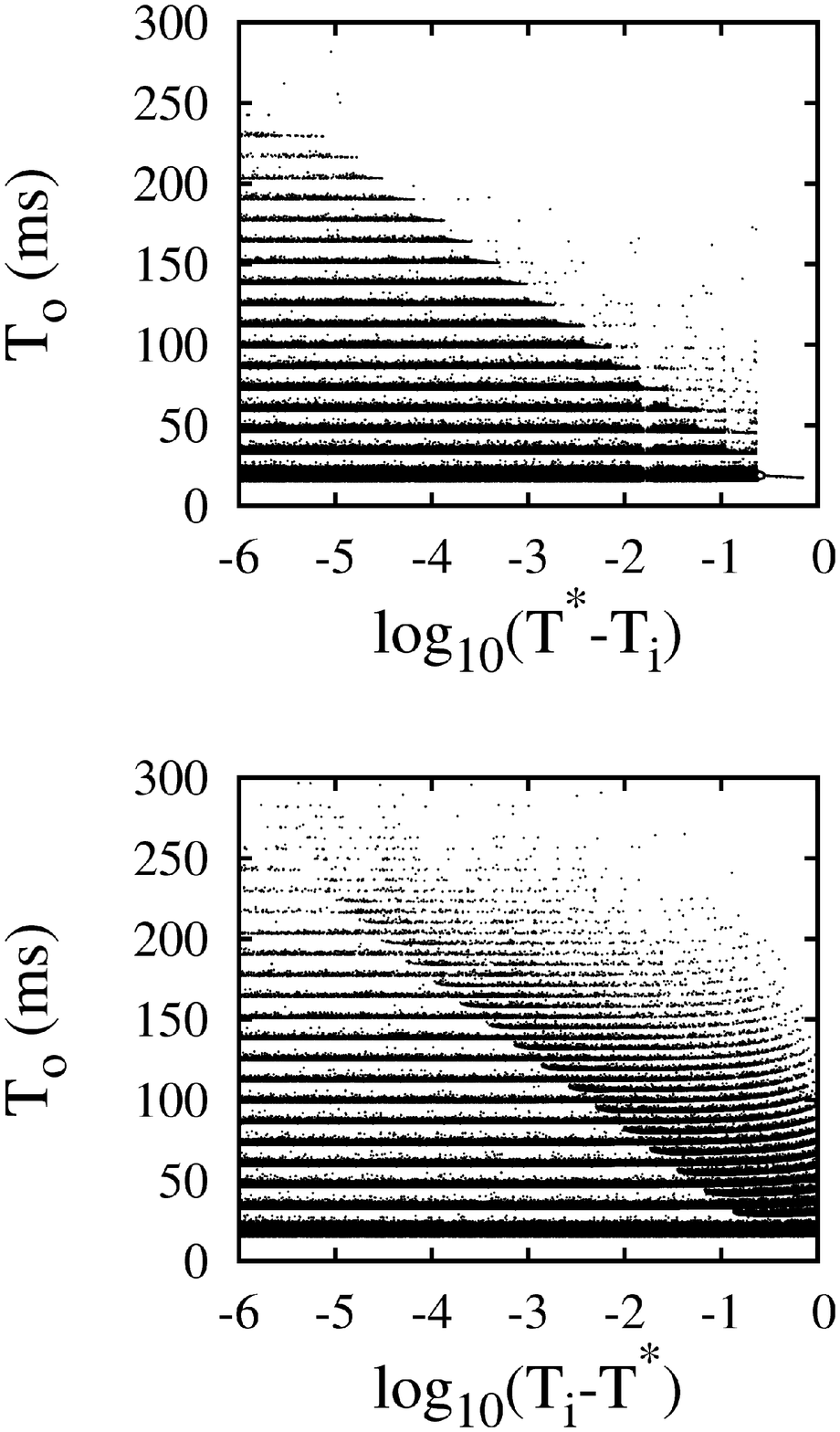}
\caption{
Scaling of the excitation edge of
(a) odd-only multiples of the input $T_i$,
and (b) all integer multiples,
in the chaotic region between $k=2$ and $k=3$.
For $g=0.2 \textrm{mS/cm}^2$, the transition occurs
at $T = 6.54175 \textrm{ms}$.
}
\label{g=0.2}
\end{figure}

\begin{figure}[ht]
\includegraphics[width=.44\textwidth]{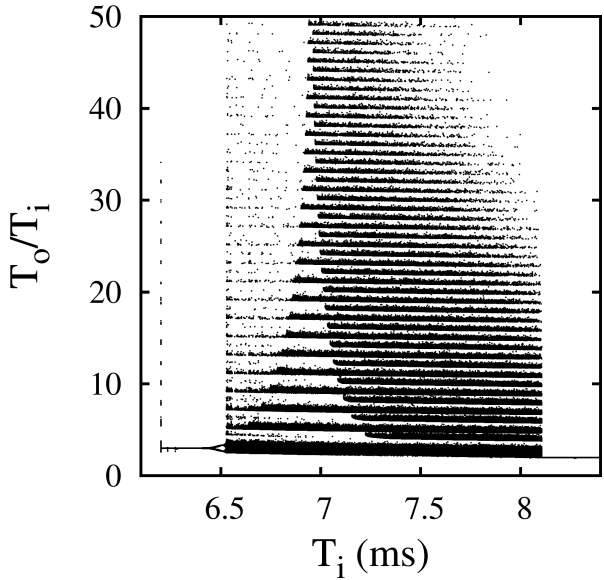}
%\vspace{-0.5cm}
\caption{
The multimodal transition at $g=0.17 \textrm{mS/cm}^2$.
}
\label{g=0.17}
\end{figure}

\begin{figure}[ht]
\includegraphics[width=.44\textwidth]{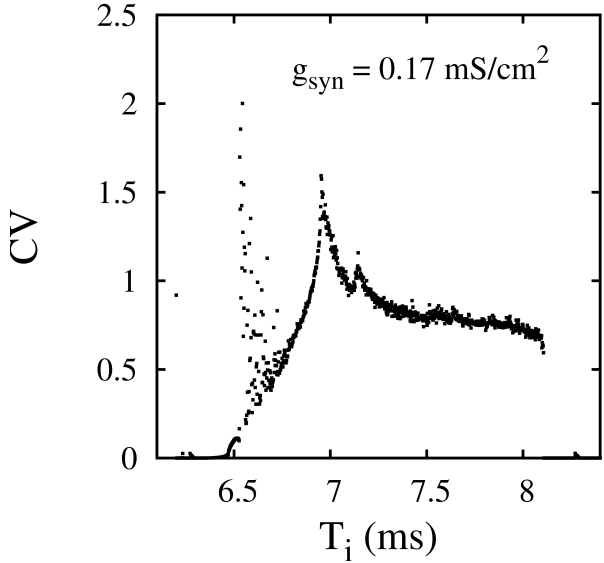}
%\vspace{-2.cm}
\caption{
Coefficient of variation for $g=0.17 \textrm{mS/cm}^2$.
The variability near $T_i = 6.6 \textrm{ms}$ is due to the proximity
to the firing threshold.
}
\label{cv_g=0.17}
\end{figure}

\begin{figure}[ht]
\includegraphics[width=.44\textwidth]{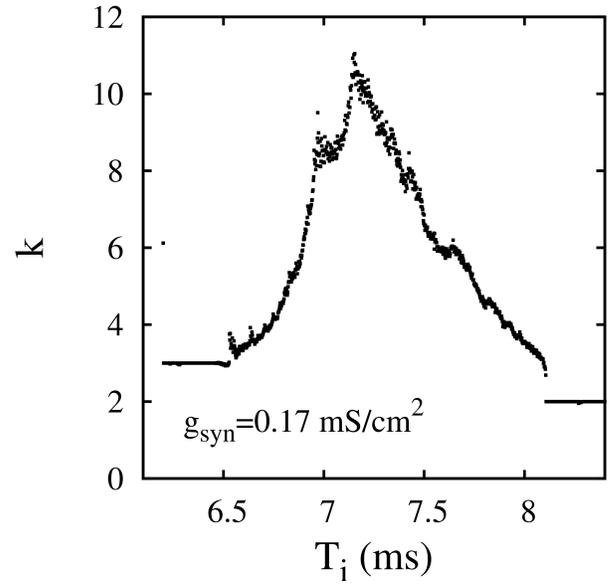}
%\vspace{-2.cm}
\caption{
The ratio $k=T_o/T_i$ for $g=0.17 \textrm{mS/cm}^2$.
The maximum of $k$ is shifted approximately 0.2 ms to the right
relative to maximum of CV (see Fig. \ref{cv_g=0.17}).
}
\label{k_g=0.17}
\end{figure}

\begin{figure}[ht]
\includegraphics[width=.44\textwidth]{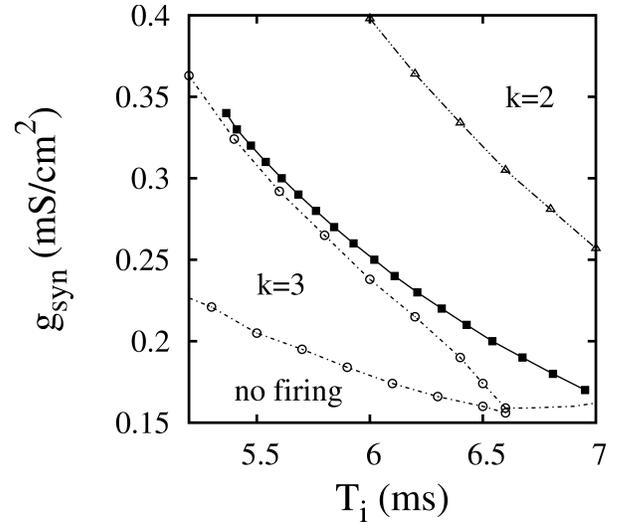}
%\vspace{-2.cm}
\caption{
The location of the multimodal transition (filled squares)
on the response diagram.
}
\label{transition}
\end{figure}

\begin{figure}[th]
\includegraphics[width=.44\textwidth]{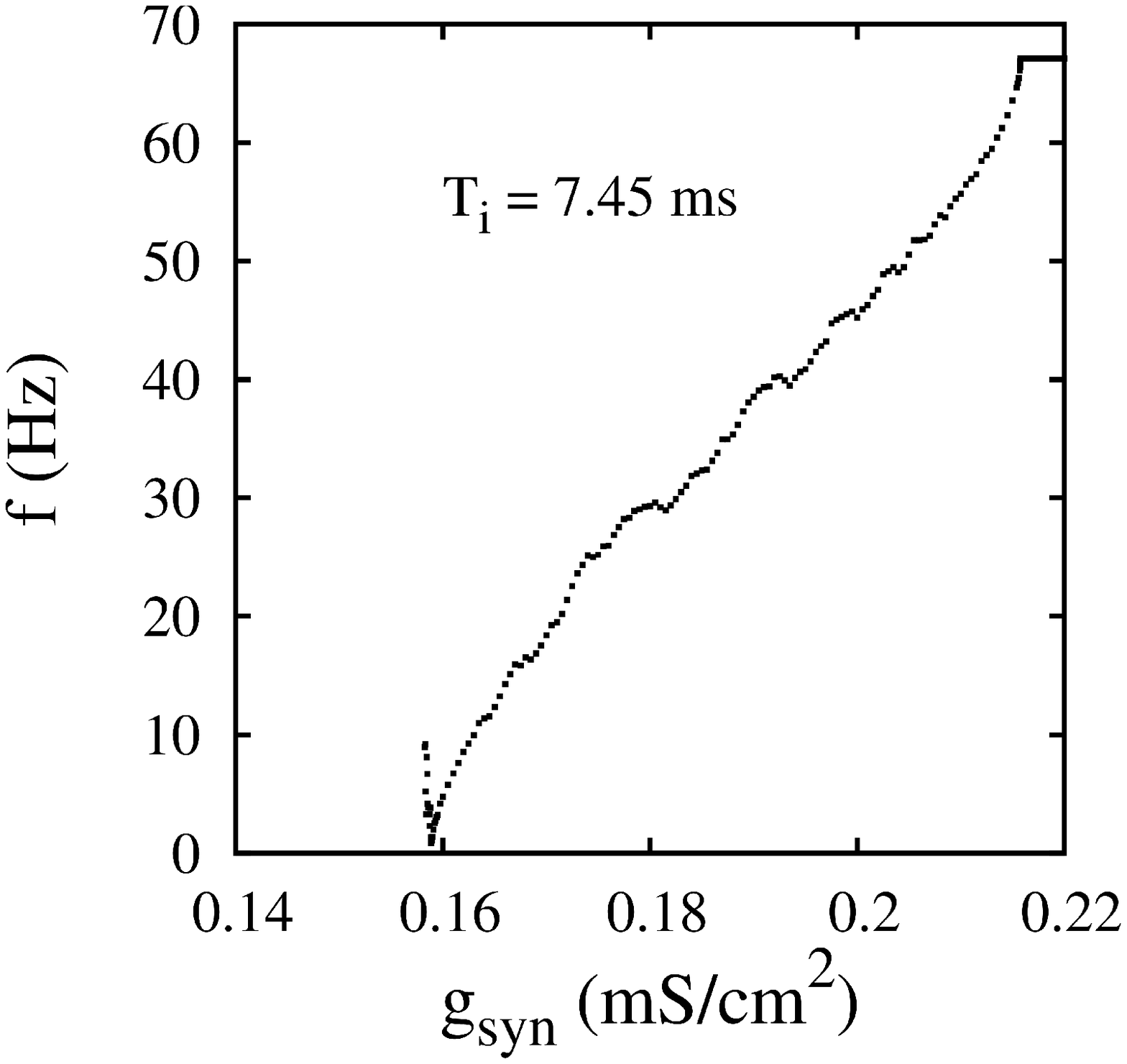}
%\vspace{-2.cm}
\caption{Frequency vs. synaptic conductivity for $T_i = 7.45 \textrm{ms}$.
Each data point is averaged over 15 runs for 60 s with different initial conditions. The initial 6 s from each run were discarded.
}
\label{T=7.45}
\end{figure}

It is interesting to note that ISI histograms (ISIH)
from some older experiments on nerve fibers
of monkeys\cite{Rose67} and single neurons in the primary
visual cortex of a cat\cite{Siegel90} show some similarity
to Fig. \ref{g04}.
Experimental histograms are sequences of diminishing peaks occuring
at integer multiples of the input interspike separation.
In Fig. \ref{g04} the lowest element of the sequence
is missing due to the refractoriness of the neuron.
Similar form of ISIH was obtained in a theoretical study of 
a bistable system stimulated by periodic
function with additive Gaussian noise,\cite{Longtin1991}
where the presence of noise was essential.
However the multimodal histogram was also obtained in a simulation
of a deterministic modification of the HH model\cite{Clay2003}.

The HH model studied here does not contain
stochastic terms. The multimodal response in Fig. \ref{g04}
is a result of a deterministic nonlinearity.
Thus noise is not the only ingredient
enabling the reproduction of the multimodal experimental ISIH.
It is possible to identify the source of multimodality
by studying ISIH in more detail.

Close to the excitation threshold, at $g_{syn} \simeq 0.2 \textrm{mS/cm}^2$,
there exists a transition from the odd-only ISIH to ISIH with all
integer multiples of $T_i$, see Fig. \ref{g=0.2}.
Near the transition the edges of high-$k$ clusters
scale logarithmically. The scaling holds
both along the $T_i$ axis and along the $g_{syn}$ axis.
It can be viewed as a competition between the odd and the even multiples
of the driving period.

A clear indication of this "spectral" transition
is the singular behavior of the coefficient of variation,
see Fig. \ref{cv_g=0.17}. At the transition CV is of order 1.
and $k$ is significantly larger than 3. The maximum $k$ occurs
approximately 0.2 ms above the singularity of CV.
One may also think of this shift as a result of relaxation
from the constraint of odd-only modes below $T^*$. At $T^*$
the highest even modes become available and this leads to the increase
of $k$.

If such transition were found experimentally
it would be a clear sign of the deterministic
nonlinear dynamics.
In the presence of noise this sharp feature would be smeared
and would vanish if noise dominates the dynamics of the system.

\section{Conclusions}

For high synaptic drives in the high frequency regime
distinguishing the action potential
from the background activity becomes problematic.
In this limit the neuron is very sensitive to small changes
of the functional form of the signal.
For periodic drive with small time constant$\tau$
and $T_i$ below $6 \textrm{ms}$ the width of the spiking
regime along the $g_{syn}$ axis scales
linearly with $T_i$. 
The quality of the neuron's response deteriorates linearly
with increasing $g_{syn}$. 
This is in contrast to findings for a sinusoidal signal,
and more generally for a class of signals satisfying
the constraint of charge balancing, where the spiking action
remains well defined in the high-frequency limit.

A mechanism of suppression of the neuron's activity
might help explain self-regulating behavior
of neocortical networks.
Various mechanisms of homeostatic action
for neural microcircuits were proposed.\cite{Muresan2007}
It would be useful to investigate whether
more realistic extensions
of the Hodgkin-Huxley model also exhibit self-regulation
in response to high-frequency inputs.
The network of such neurons would have
a "safety switch" built in at the level of individual cells.
For $T_i$ between 4 and 6 ms
the upper critical synaptic conductivity is of order
$2 \textrm{mS/cm}^2$, which is in the realistic range
for neocortical pyramidal neurons\cite{Ho2000}.

The input ISI of $4-8 \textrm{ms}$ is important to understanding
the dynamics of the Hodgkin-Huxley model.
In the chaotic region between the $k=2$ and $k=3$ locked states
the coefficient of variation of ISI has a singularity
at the transition between the odd-only
and all-integer multiples of the driving period.
The odd modes dominate in the vicinity of the $k=3$ state.
The low-$k$ (high-frequency) bands vanish logarithmically near
the line of critical points ($g_{syn}$,$T^*$).
The firing rate has a minimum at $T_i \simeq T^* + 0.2 \textrm{ms}$.
Periodically stimulated giant axons of squid have similar nonmonotonic
dependence of the firing rate on the current pulse amplitude
between the $k=2$ and $k=3$ states\cite{Takahashi1990}.
This experiment also showed linear dependence of the firing rate
on pulse amplitude near the threshold for $T_i > T^*$, similarly
to Fig. \ref{T=7.45}. Although the experimental pulses were rectangular,
different from the $\alpha(t)$ form with an exponential tail,
the qualitative features do not depend much on the precise shape
of a pulse. For short pulses the neuron's reaction
is determined mainly by the time integral
of the stimulus.

The multimodal response occuring in certain
sensory neurons may result from noise\cite{Longtin1991}
or deterministic nonlinearity\cite{Kaplan1996}. 
It would be interesting to look for experimental evidence
of the odd-all transition. It found, it would be a clear evidence that
the neuron dynamics is dominated by nonlinearity, not noise.

The behavior of the model at small $T_i$ may be useful
to both coincidence detection and estimation of the signal strength.
The optimal sensitivity in this case is inversely proportional
to frequency.

Our calculation also supports the view
expressed by authors of Ref. \cite{Guckenheimer2002}
that boundaries between various parts of the response
diagram are not always clear-cut and may form
complicated patterns. This statement
also applies to the excitation threshold
in the chaotic regime.

In the Hodgkin's classification of intrinsic excitability\cite{Hodgkin1948}
class 1 neurons maintain firing at arbitrarily low frequencies in response
to weak inputs and have continuous frequency-current ($f$-$I$) curve.
Class 2 neurons fire with certain
relatively large frequency, usually of order 40-50 Hz,
when stimulus exceeds threshold and have a discontinuous
$f$-$I$ curve.
Class 1 and class 2 neurons sometimes are described as integrators
and resonators respectively\cite{Izhikevich2000}.
According to the commonly held view
a neuron cannot be an integrator and resonator at the same time. 
However we showed that the deterministic HH neuron 
in a chaotic regime near excitation threshold
may oscillate with arbitrarily small frequencies
and may perform integration at time scales much longer than
the period of its main resonance.
The character of the response depends strongly
on the functional form of the stimulus and parameters of the model.
A recent study showed that the same pyramidal neurons
behave as integrators in vitro and resonators in vivo.\cite{Prescott2008}

The multimodal response of the HH neuron near 140-180 Hz
is not a typical resonance since no particular frequency
is preferred. The multiples of the driving frequency
alternate chaotically. The average output frequency
depends nonmonotonically on the stimulus amplitude.
Similar nonmonotonic $f$ vs. $I$ relation was found
in periodically stimulated giant axons of squid\cite{Takahashi1990}.
Smaller stimuli favor higher multiples of the driving period.
Studies of large neuronal networks
of various types suggest that there may be a complex
interplay between the integrating behavior and the resonant
action.\cite{Muresan2007}

The ability to precisely control the nerve cell's potential
oscillations is important in constructing devices
performing the procedure known as Deep Brain Stimulation\cite{Benabid1991,Gross2000,McIntyre2004},
which operate at frequencies above 100 Hz.
While our model does not satisfy the charge-balancing constraint
required in the stimulation of in-vivo systems, we believe the present
study improves our understanding 
of high-frequency neural oscillators.

\acknowledgments
The author thanks J. W. Mozrzymas, D. W\'ojcik, K. Bodova,
T. Burwick, and P. Suffczy\'nski for discussions.

Computations were performed in the Computer Center
of the Tri-city Academic Computer Network in Gdansk.

%\floatfix
%\section*{References}
%\begin{thebibliography}{9}
\bibliography{h2009}

\end{document}